\documentclass[a4paper,12pt, times]{article}

\usepackage{amsmath}
\usepackage{authblk}
\usepackage{geometry}
\usepackage{graphicx}
\usepackage{subcaption}
\usepackage{subfloat} 
\usepackage{tabularx}
\usepackage{makecell}
\usepackage{multirow}
\usepackage{changepage}
\usepackage{setspace}
\usepackage{pifont}
\usepackage{enumitem}
\usepackage{soul}
\usepackage{xcolor}
\newlist{tabitem}{itemize}{1}
\setlist[tabitem]{wide=0pt, nosep, leftmargin= * ,label=\textbullet,after=\vspace{-\baselineskip},before=\vspace{-0.6\baselineskip}}

\providecommand{\keywords}[1]{\textbf{\textit{Keywords---}} #1}

\begin{document}
\title{An Algorithm for Modelling Escalator Fixed Loss Energy for PHM and sustainable energy usage}
\author[1]{HU, Xuwen}
\author[1]{QIU, Jiaqi}
\author[1]{LIN, Yu}
\author[1,2]{ZWETSLOOT, Inez Maria}
\author[3]{LEE, William Ka Fai}
\author[3]{YEUNG, Edmond Yin San}
\author[3]{YEUNG, Colman Yiu Wah}
\author[3]{WONG, Chris Chun Long}
\affil[1]{Department of Systems Engineering, City University of Hong Kong}
\affil[2]{Department of Business Analytics, University of Amsterdam}
\affil[3]{MTR Corporation Limited}
\date{}
\maketitle

\begin{abstract}
Prognostic Health Management (PHM) is designed to assess and monitor the health status of systems, anticipate the onset of potential failure, and prevent unplanned downtime. In recent decades, collecting massive amounts of real-time sensor data enabled condition monitoring (CM) and consequently, detection of abnormalities to support maintenance decision-making. Additionally, the utilization of PHM techniques can support energy sustainability efforts by optimizing energy usage and identifying opportunities for energy-saving measures. Escalators are efficient machines for transporting people and goods, and measuring energy consumption in time can facilitate PHM of escalators. Fixed loss energy, or no-load energy, of escalators denotes the energy consumption by an unloaded escalator. Fixed loss energy varies over time indicating varying operating conditions. In this paper, we propose to use escalators' fixed loss energy for PHM. We propose an approach to compute daily fixed loss energy based on energy consumption sensor data. The proposed approach is validated using a set of experimental data. The advantages and disadvantages of each approach are also presented, and recommendations are given. Finally, to illustrate PHM, we set up an EWMA chart for monitoring the fixed loss over time and demonstrate the potential in reducing energy costs associated with escalator operation. 
\end{abstract}


\newpage
\section{Introduction}

Nowadays, massive amounts of real-time sensor data are automatically collected. This has enabled prognostics and health management (PHM) as an important solution for system reliability and safety. For details on PHM, the reader is referred to \cite{Tsui15}. PHM is applied in many different fields. For example, aerospace engineering \cite{Batzel09}, mechanical engineering \cite{Che19}, mechanical design \cite{Hu22}, maintenance policy \cite{Jonge17}\cite{Dimitris22} and condition-based maintenance \cite{Zheng21}\cite{Liu22}. For our project we are focussing on PHM for escalators.

Escalators, a complex mechanical system for vertical transportation, play a critical role in our daily life. Escalators have been studied in various fields such as analysis of accidents \cite{Xing20}, energy consumption \cite{Al-sharif11}\cite{Uimonen17} and energy efficiency \cite{De12}\cite{Uimonen20}. Besides that, escalator PHM is another important topic, with studies mainly concentrating on vibration of escalator components\cite{Elasha14}\cite{Huo20}. We believe that escalator energy consumption can also be used to assess the health status of an escalator. 

As suggested by Al-Sharif \cite{Al-sharif11}, the energy consumption of an escalator can be divided into two parts: variable and fixed loss. Variables loss energy is the changing energy consumption used for regenerated from transporting passengers. Fixed loss energy denotes the unloaded escalator energy consumption, which is used to overcome the friction between steps, escalator belt, motor and gearbox \cite{Ma09}. Vertical rise and mechanical design are important factors influencing the fixed loss. Also, the fixed loss is affected by the thermal process in the handrail, gearbox etc. \cite{Carrillo13}. Fixed loss varies over time and is related to the operating condition of the escalator. Therefore, we propose to use fixed energy loss for the PHM process of escalators in this study. 

Studies for escalator PHM using energy consumption are rare.  The reason is that a direct and accurate measurement of the escalator energy consumption is usually unavailable as most escalators are not equipped with a connected energy meter \cite{Uimonen17}. De Almeida \cite{De12} stated that the major limitation to energy efficiency is the lack of energy consumption monitoring.


In this project, long-term continuous energy consumption sensor data for the escalator are collected. This allows PHM for escalators by monitoring fixed loss, which leads to reductions in energy consumption, minimize wasted energy, and ultimately achieve substantial energy savings throughout the escalator's operational lifetime. We propose a new algorithm to compute time-varying fixed loss based on total energy consumption, recorded every minute for more than one year. In addition, we perform a validation study to test the accuracy of our approach. In this experiment, several escalators are selected to run for one hour without passengers to obtain an exact value for the fixed loss. These experimental fixed loss values are compared with two existing methods and our proposed methodology. Our approach shows broad applicability to various cases of energy consumption patterns and accurate performance. In addition, we also illustrate how the time-varying fixed loss values can be used for PHM by implementing a monitoring method to detect changes in the fixed loss and facilitate maintenance decisions.

Summarizing the contribution of this work is that 
\begin{itemize}
    \item[\ding{108}] We propose a novel algorithm to model time varying fixed loss
    \item[\ding{108}] And we perform experiments to tune and verify the algorithm
\end{itemize}
This proposed algorithm can be used for PHM of escalators as well as to enable energy efficient escalator operations. 

The rest of the paper is organized as follows. Section 2 presents the data description. In Section 3, we introduce our proposed algorithm as well as two existing methods for estimating time-varying fixed loss. Section 4 discusses parameter tuning and comparisons among fine-tuned models. Section 5 illustrates using time-varying fixed loss for PHM by setting up a monitoring methodology and implementing it on a test escalator. Some concluding remarks are given in Section 6.

\section{Energy Data Description}
The fixed loss energy is illustrated in this section. Using energy meters, we collected energy consumption data from 20\footnote{The original study includes 24 escalators but for our purpose we omit escalators that are bi-directional or in operation for less than 2 hours per day} escalators installed in several public underground stations in Hong Kong from November 2020 to May 2022. There are ten upward-escalators and ten downward-running escalators in our samples. The total energy consumption is recorded every minute, 24 hours per day (1440 records per day). Most escalators start service at 5:30 am and turn off around 1 am. Therefore, we define every 4 am to the following 4 am as one working day. 

We define the day $t$, the minute $i$ of the day, the energy is measure in Watt-hour, total energy consumption $E_{t,i}$ at every minute, and fixed loss energy $F_t$ for one day. The daily energy consumption of day $t$ is a set $ E_t=\left\{{E_{t,i}} \right\}$, where $i=1,2,...,n$. The escalator turns on at $i=1$ and turns off at $i=n$, hence $n<1440$. 

\begin{figure}
	\centering
	\begin{subfigure}[b]{0.48\textwidth}
		\centering
		\includegraphics[scale=0.26]{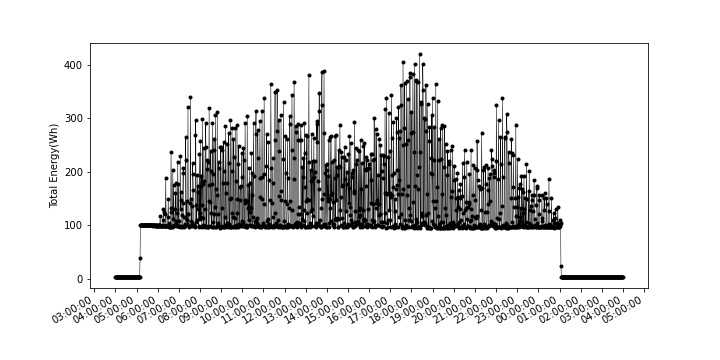}
		\caption{Escalator \#1}
		\label{fig:scatter_passenger_up}
	\end{subfigure}
	\hfill
	\begin{subfigure}[b]{0.48\textwidth}
		\centering
		\includegraphics[scale=0.26]{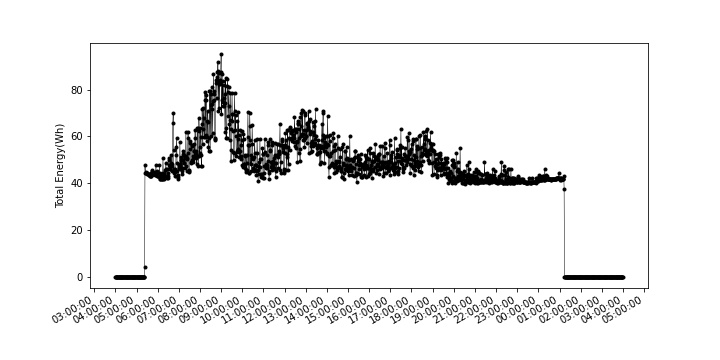}
		\caption{Escalator \#3}
		\label{fig:rushhour_passenger_up}
	\end{subfigure}
	\caption{Total Energy consumption for Upward escalator in one working day}
	\label{fig:upward_escalators}
\end{figure}

\begin{figure}
	\centering
	\begin{subfigure}[b]{0.48\textwidth}
		\centering
		\includegraphics[scale=0.26]{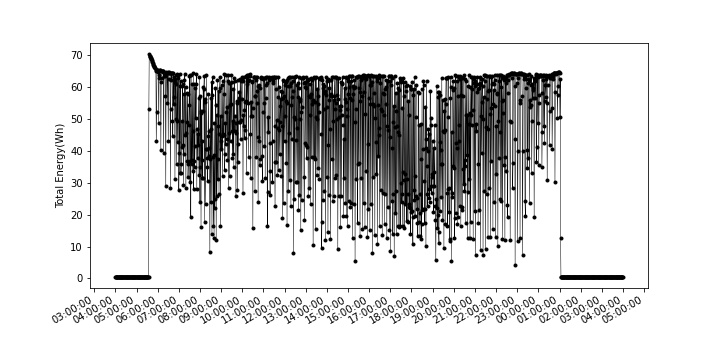}
		\caption{Escalator \#7}
		\label{fig:scatter_passenger_down}
	\end{subfigure}
	\hfill
	\begin{subfigure}[b]{0.48\textwidth}
		\centering
		\includegraphics[scale=0.26]{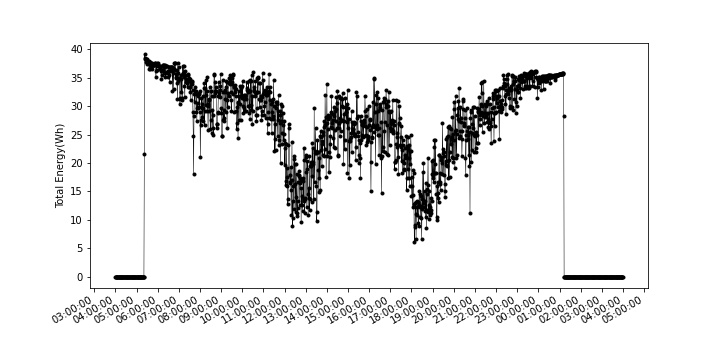}
		\caption{Escalator \#5}
		\label{fig:rushhour_passenger_down}
	\end{subfigure}
	\caption{Total Energy consumption for Downward escalators in one working day}
	\label{fig:downward_escalators}
\end{figure}

Energy consumption consists of fixed-loss and variable loss energy,  which is influenced by the direction of escalators \cite{Al-sharif11}. For the upward escalator, the total energy is the sum of fixed loss $E_{t,i}$ and variable energy consumption $V_{t,i}$ caused by passengers, which is $E_{t,i }= F_{t} + V_{t,i}$. Conversely, fixed loss minus variable loss is considered total energy consumption for the downward escalator due to gravity, written as $E_{t,i }= F_{t} - V_{t,i}$. Figures \ref{fig:upward_escalators} and \ref{fig:downward_escalators} show examples of different escalators for both directions. Every data point represents the total energy consumption at that minute. As shown in Figures \ref{fig:rushhour_passenger_up} and \ref{fig:rushhour_passenger_down}, it is evident that there are several waves of passengers mostly, around rush hours and noon. In contrast, in Figures \ref{fig:scatter_passenger_up} and \ref{fig:scatter_passenger_down}, the passenger using pattern is scattered. 

Close inspection of Figure 2 shows that at the start of operation the escalator
is using higher levels of energy for operation which then gradually drop to the fixed loss energy level after about 30 to 60 minutes. This phenomenon is quite well-known and is caused by the thermal processes within the system \cite{Carrillo13}. This effect can be observed in all escalators when they start up, though the size of the effect varies which makes it less obvious to be spotted, for example, in Figure \ref{fig:upward_escalators}.

\section{Methodology}
In this section,  the proposed approach to computing the fixed loss energy from sensor data is introduced, which we refer to as the "Optimization approach". Two methods from literature \cite{Al-sharif11} are also presented, which we refer to as the "Classical approach" and the "Engineering approach". 
    \subsection{Proposed Method: Optimization approach}
    The objective is to identify an energy value that has the maximum proximity to a significant number of data points while simultaneously being as small (or large) as possible when compared to all other data points. In other words, finding the fixed loss can be formulated as an optimization problem. 
    
\begin{figure}[h!]
    	\centering
    	\begin{subfigure}[b]{0.48\textwidth}
    		\centering
    		\includegraphics[scale=0.35]{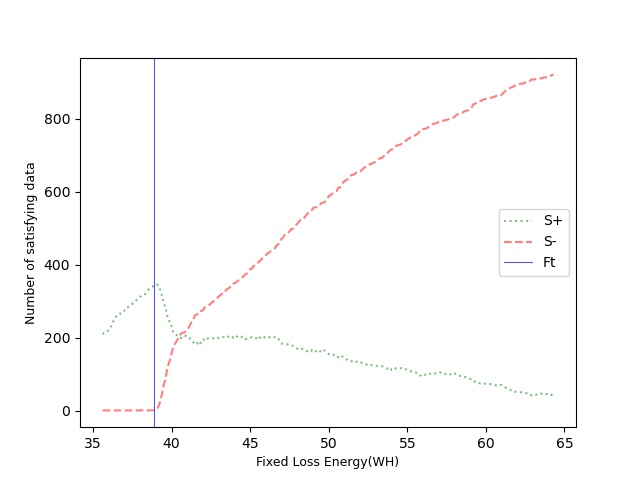}
    		\caption{$\lvert$$S^+$$\rvert$ and $\lvert$$S^-$$\rvert$}
    		\label{fig:optimization for upward escalator}
    	\end{subfigure}
    	\hfill
    	\begin{subfigure}[b]{0.48\textwidth}
    		\centering
    		\includegraphics[scale=0.26]{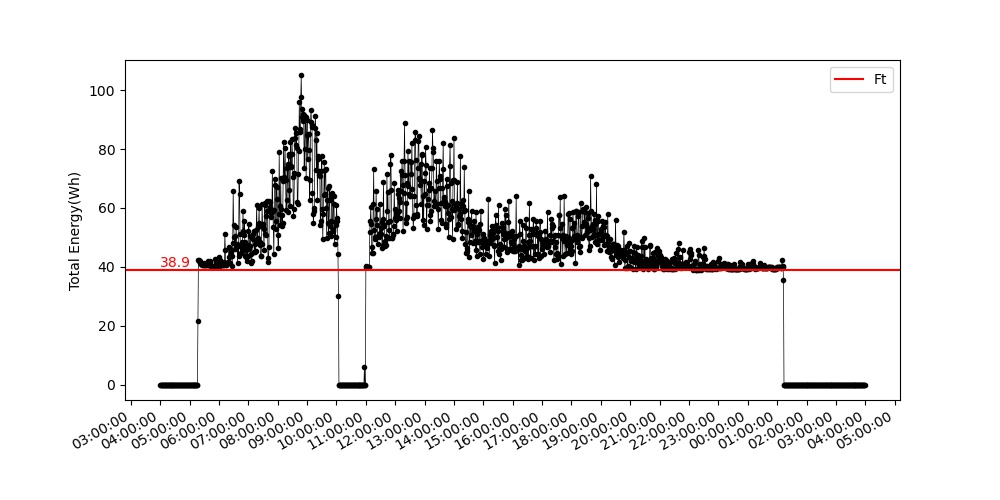}
    		\caption{Optimal $F_t$ and $E_t$}
    		\label{fig:Energy with fixed loss-upward}
    	\end{subfigure}
    	\caption{Selecting $F_t$ with for Upward escalator using Optimization Approach (Escalator \#3)}
    	\label{fig:optimization_upward_escalators}
    \end{figure}

\begin{figure}[h!]
    	\centering
    	\begin{subfigure}[b]{0.48\textwidth}
    		\centering
    		\includegraphics[scale=0.35]{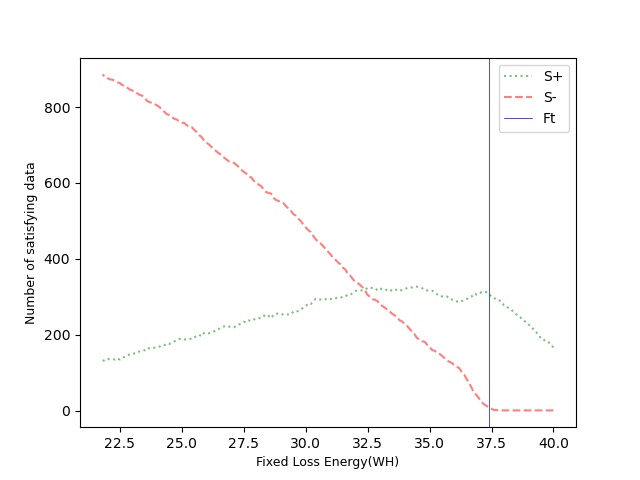}
    		\caption{$\lvert$$S^+$$\rvert$ and $\lvert$$S^-$$\rvert$}
    		\label{fig:optimization for downward escalator}
    	\end{subfigure}
    	\hfill
    	\begin{subfigure}[b]{0.48\textwidth}
    		\centering
    		\includegraphics[scale=0.26]{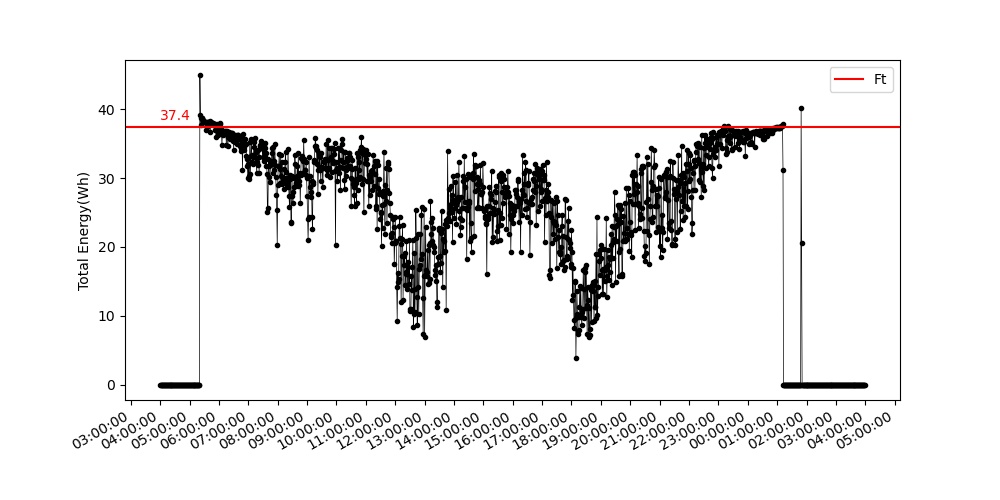}
    		\caption{Optimal $F_t$ and $E_t$}
    		\label{fig:Energy with fixed loss-downward}
    	\end{subfigure}
    	\caption{Selecting $F_t$ with for Downward escalator using Optimization Approach (Escalator \#4)}
    	\label{fig:optimization_downward_escalators}
    \end{figure}

    For the upward escalator (e.g., Figure \ref{fig:upward_escalators}), define the set $ \ S_t^ +$ as those minutes $i$ that energy consumption $E_{t,i}$ is close to but larger than $F_t$. Similarly, define the set $\ S_t^-$ as those minutes $i$ that energy consumption $E_{t,i}$ is close to but smaller than $F_t$; for some pre-specified fixed loss energy $F_t$.
    \begin{equation}
        \begin{aligned}
            \ S_t^+ &= \{i \quad {\rm s.t.} \quad 0<E_{t,i} - F_t <\delta\} \\
            \ S_t ^- &= \{i \quad {\rm s.t.} \quad 0<F_t - E_{t,i}<\delta\}
        \end{aligned}
    \end{equation}
    Conversely, for the downward escalator, define the set $ \ S_t^ +$ as those minutes $i$ that energy consumption $E_{t,i}$ is close to but smaller than $F_t$. Similarly, define the set $\ S_t^-$ as those minutes $i$ that energy consumption $E_{t,i}$ is larger than $F_t$;  for some pre-specified fixed loss energy $F_t$. 
    \begin{equation}
        \begin{aligned}
            \ S^+ = \{i \quad {\rm s.t.} \quad 0<  F_t - E_{t,i} <\delta\} \\
            \ S ^- = \{i \quad {\rm s.t.} \quad  0< E_{t,i} - F_t <\delta\}
        \end{aligned}
    \end{equation}
    Now we wish to set $F_t$ such that $\ S_t^+$ is a set as large as possible and set $\ S_t^-$ is a relatively small set, i.e., the $F_t$ is a horizontal line that lies close to the bottom (top) of the energy profile for upward (downward) escalators. Hence, the process of finding the optimal $F_t$ can be expressed as
    \begin{equation}
        \begin{aligned}
            \text{maximize} & \quad \lvert \ S_t^+\lvert - \lvert \ S_t^-\lvert \\
            \text{subject to}  &  \quad  F_{t} >0
        \end{aligned}
    \end{equation}
    Figures \ref{fig:optimization_upward_escalators} and \ref{fig:optimization_downward_escalators} show the process of locating the optimal fixed loss $F_t$ for a upward (Figure \ref{fig:optimization_upward_escalators}) and downward (Figure \ref{fig:optimization_downward_escalators}) escalators. Figures \ref{fig:optimization for upward escalator} and \ref{fig:optimization for downward escalator} illustrate the difference between $\lvert$$S^+$$\rvert$ and $\lvert$$S^-$$\rvert$. The values of the fixed loss are visualised in Figures \ref{fig:Energy with fixed loss-upward} and \ref{fig:Energy with fixed loss-downward}. It is shown that the estimated fixed loss $F_t$ fits the energy consumption perfectly. The selection of the tuning parameter $\delta$ and the validation of our method will be discussed in Section 4.2. In the following subsection, we firstly introduce two existing alternative approaches to computing $F_t$. 
    
    \subsection{Existing Methods}
        \subsubsection{Classical Approach}
            The classical approach was originally proposed by Al-Sharif \cite{Al-sharif11},  and also discussed in other studies \cite{Al-sharif98} \cite{Kuutti13}. The approach estimates the fixed loss by computing the average total energy consumption of the last half hour of operating time. This is built on the assumption that little to no passengers use the escalators in the last 30 minutes of operating time. 
            The fixed loss energy can be computed as 
            \begin{equation}
                \begin{aligned}
                \ F_t &= 
                \frac{\sum_{i=n-29}^{n} E_{t,i}}{30} \
                \end{aligned}
            \end{equation}
        \subsubsection{Engineering Approach}
            From an engineering perspective, fixed loss energy presents the vacant status of a running escalator. And it equals, the minimum value of total energy consumption for an upward escalator or the maximum value for the downward escalator. In this approach we use this easy heuristic. This method is not directly taken from any papers but rather it is based on what us used in practice. 
            
            The first operating hour is deleted to avoid inaccurate and biased energy consumption due to the thermal process. Therefore, the daily energy consumption set becomes $E_t = \left\{{E_{t,i}} \right\}$ where $i=61,…,n$. We select, for upward escalators, the $p\%$ smallest values in $E_t$, i.e., $\hat{E_t^p}$. For downward escalators, we select the $p\%$ largest value in $E_t$, i.e., $\hat{E_t^p}$.
            
            Then, the fixed loss energy can be calculated as the median of $\{\hat{E}^p_{t,i}\}$
            \begin{equation}
                \begin{aligned}
                \ F_t &= {\text{Median}}\left\{\hat{E}^p_{t,i}\right\}\
                \end{aligned}
            \end{equation}
            In other words, non-consecutive $p\%$ of processed data $E_{t,i}$ will be taken, as we expect escalators are vacant at varying time points during the day. We take the median to stay away from dominated outliers, and to smooth some of the noises. The selection of tuning parameters $p$ will be discussed in Section 4.2.

\section{Comparison Analysis}
    In this section, the performances of the three methods are compared by an experiment. The experimented data are introduced in Section 4.1. The selection of tuning parameters ($\delta$ for the optimization approach and $p$ for the engineering approach) are discussed in Section 4.2. Then, in Section 4.3, tuned models are applied to the experimental data to get the fixed loss and make a comparison of the accuracy of the three approaches. In the final subsection, the comparison results are discussed.
    \subsection{Experiment data}
    In Hong Kong, it is challenging to obtain fixed loss energy in daily operations, as there are always passengers using the escalator. 
    To obtain an accurate label for validation of fixed loss energy estimation, we conducted 11 experiments\footnote{Actually, we conducted a total of 18 experiments on 8 different escalators. However, we chose to exclude 6 of these datasets as they show unstable energy patterns influenced by thermal processes in the system, which could lead to biased fixed loss estimates.}, five of which were conducted with an the upward escalator and six for a downward one. 
    
\begin{figure}
    	\centering
    	\begin{subfigure}[b]{0.48\textwidth}
    		\centering
    		\includegraphics[scale=0.26]{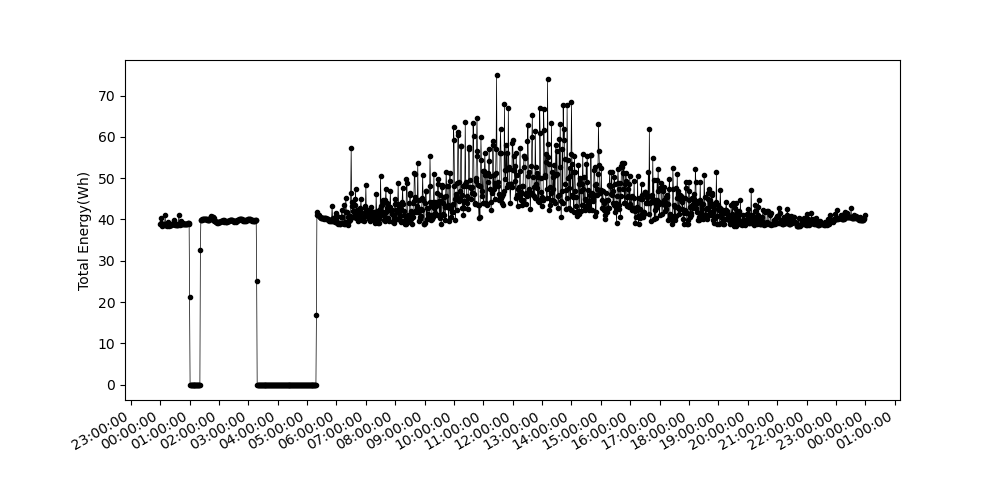}
    		\caption{Escalator \#3}
    		\label{fig:experiment for upward escalator}
    	\end{subfigure}
    	\hfill
    	\begin{subfigure}[b]{0.48\textwidth}
    		\centering
    		\includegraphics[scale=0.26]{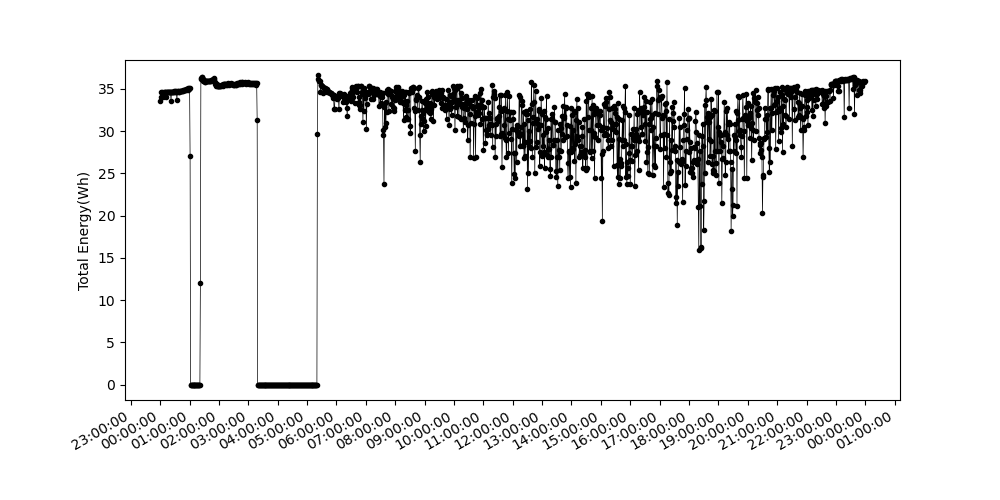}
    		\caption{Escalator \#4}
    		\label{fig:experiment for downward escalator}
    	\end{subfigure}
    	\caption{Energy Consumption during experiment day}
    	\label{fig:experiment}
    \end{figure}

    All experiments were conducted during the station non-service time and lasted for approximately one hour. 
    This ensures that the recorded energy consumption values represent energy consumed by the escalator running without passengers. The experimental data are then used to optimise parameters, taking both the mean and standard deviation of the estimation errors into consideration. Figure \ref{fig:experiment} provides a visual representation of two of the experiments, which were conducted between 1:30 am to 3:00 am.

    During the experiments, it was observed that the energy consumption was unstable at the beginning due to the thermal process. These energy readings are higher than the actual fixed loss.
    To get an accurate label of fixed loss, we need to determine when the thermal process is finished and the fixed loss becomes stable. As the actual readings fluctuate, the 5-minute moving average is applied to smooth the data. Once the moving average converges, we deem the energy consumption stable. Therefore, the converged moving average energy consumption $F_t^E$ is considered the actual fixed loss of that day, which we use as the label for the validation and tuning of our proposed method.
    
    \subsection{Fine-tuning}
    There are 11 experimental fixed losses($F_t^{E}$) considered as the ground truth (reported in Table \ref{T1}). They are used to tune parameters $\delta$ and $p$, such that the fixed loss computed by the two approaches is as close as possible to ($F_t^{E}$) with minimum mean and variability of estimation errors across the 11 experiments.
    
    \begin{figure}
    	\centering
    	\begin{subfigure}[b]{0.48\textwidth}
    		\centering
    		\includegraphics[scale=0.26]{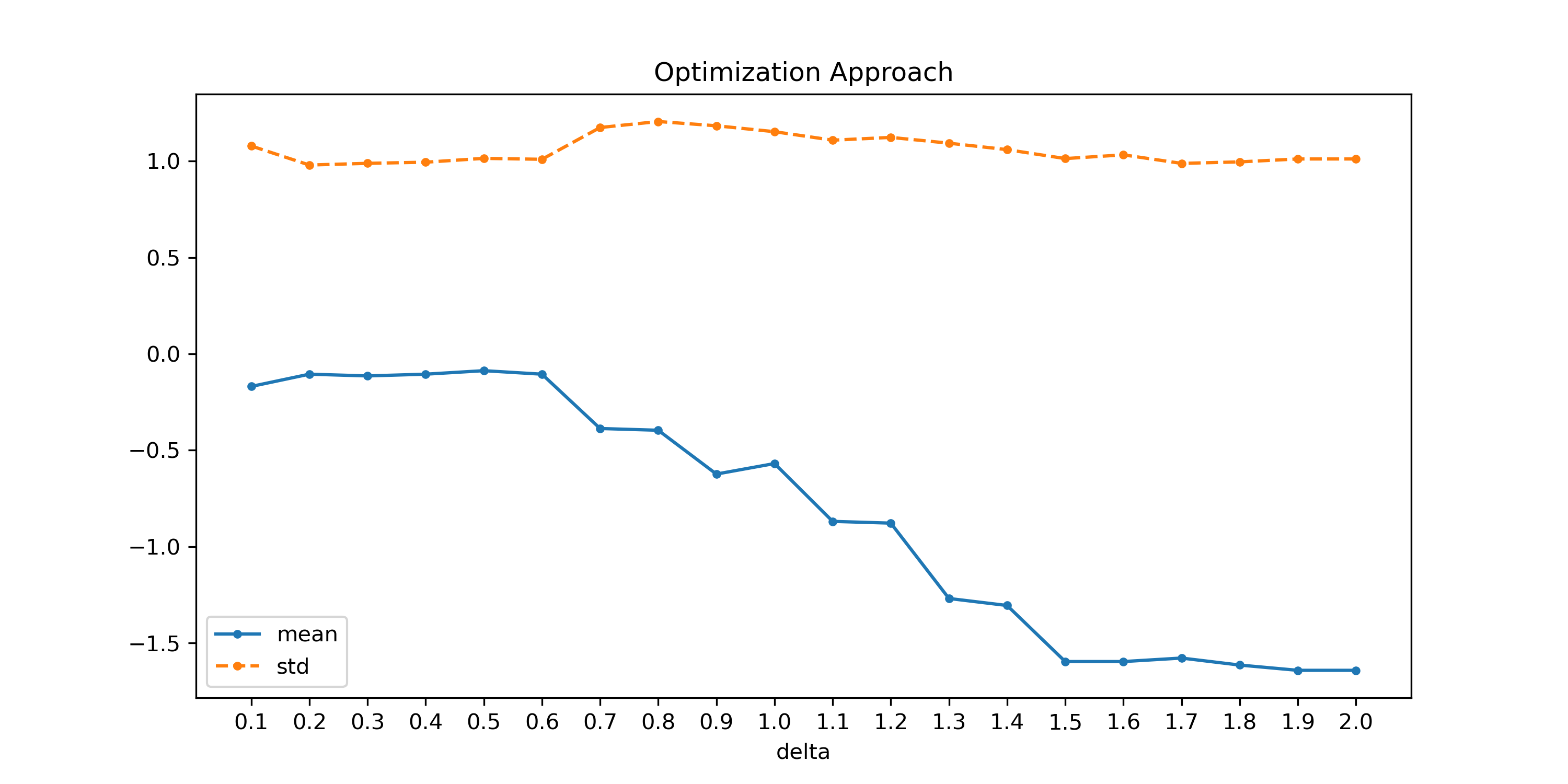}
    		\caption{Optimization approach}
    		\label{fig:optimization tuning}
    	\end{subfigure}
    	\hfill
    	\begin{subfigure}[b]{0.48\textwidth}
    		\centering
    		\includegraphics[scale=0.26]{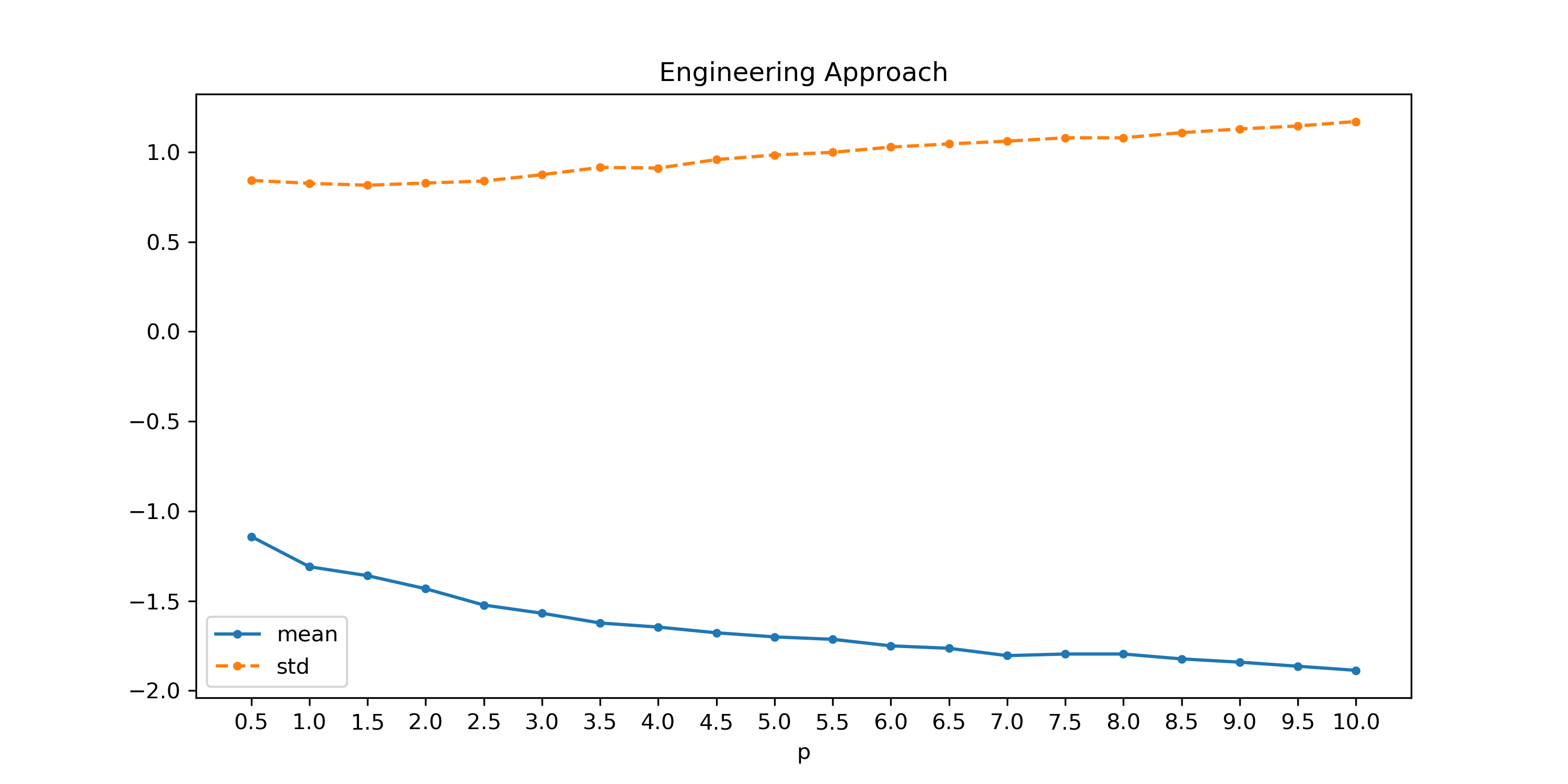}
    		\caption{Engineering approach}
    		\label{fig:engineering tuning}
    	\end{subfigure}
    	\caption{Tuning parameters in optimization approach and engineering approach with corresponding mean error and standard deviation}
    	\label{fig:tuning}
    \end{figure}
    
    The mean and standard deviation for the optimization approach estimation errors are presented in Figure \ref{fig:optimization tuning}. It is observed that the mean and standard deviation increased as the threshold $\delta$ increased beyond 0.5. Conversely, when $\delta$ is smaller than 0.5, the mean approached 0. Moreover, the standard deviation and mean are significantly larger for a threshold of $\delta=0.1$ than for thresholds of $\delta \in (0.3, 0.5)$. Therefore, a threshold of 0.3 was chosen as the optimal threshold due to its smaller mean error and standard deviation.
    
    Figure \ref{fig:engineering tuning} illustrates the changes in the mean and standard deviation of the estimation errors as a function of $p$ for the engineering approach. It can be seen that there is a slight increase in the mean as the value of $p$ increases, and consistently yields an underestimated of about 1. Meanwhile, the standard deviation increased slightly as the parameter $p$ increased. In other words, the Engineering approach is slightly biased. In addition, the values of $p$ determines the number of data points in the set $E_t^{p}$. In order to have a robust fixed loss estimate we recommend that this set is not too small hence $p$ should not be small either. Therefore, we recommend to set $p=5\%$. Though this is not the smallest variability observed in Figure \ref{fig:tuning}, it does ensure a robust fixed loss estimate.

    \subsection{Theoretical Comparison}
    Next, we compare the accuracy of the three approaches using the experimental and observation energy data. In Table \ref{T1}, we present the fixed loss value based on the experiment as well as the $F_t$ computed with our 3 methods on experiment day during normal working hour. Table \ref{T1} also presents the corresponding absolute mean error and error percentage. The results show that the optimization approach produces the most precise estimate of the fixed loss energy value with the lowest mean error ($\tau=-0.11$), whereas the classical approach provides the worst estimation ($\tau=2.87$). Interestingly, the estimated fixed loss energy for upward escalators is lower than the experimental value, while for downward escalators, it is higher than the actual value. In contrast, the engineering approach systematically underestimates the fixed loss energy, indicating a bias in this approach.
    
    \begin{table*}
        \resizebox{\columnwidth}{!}{
         \begin{tabular}{|c|l|ccc|ccc|ccc|}
            \hline
            ID&
            \multirow{2}*{Experiment} &
            \multicolumn{3}{c|}{Classical Approach} & \multicolumn{3}{c|}{Engineering Approach($5\%$)}
        & \multicolumn{3}{c|}{Optimization Approach($0.3$)}\\
           \quad&Fixed Loss& Fixed Loss& Error $\tau$& $\tau$\%& Fixed Loss& Error $\tau$ & $\tau$\%& Fixed Loss& Error $\tau$ & $\tau$\%\\ \hline
        \#12 & 51.54  * & 51.1 & -0.44 & 0.9\%& 50.3 & -1.24 &-2.4\%& 50.5 & -1.04 &-2.0\% \\ 
        \#12 & 51.12  * & 49.27 & -1.85 & -3.6\% & 49.4 & -1.72 &-3.4\% & 49.6 & -1.52 & -3.0\%\\
        \#20 & 68.82 & 62.85 & -5.97 & -8.7\% & 66.2 & -2.62 & -3.8\% &69.3 & 0.48 & 0.7\%\\
        \#20 & 67.06 & 60.75 & -6.31 &-9.4\% & 65.2 & -1.86 &-2.8\%& 68 & 0.94 &1.4\%\\
        \#20 & 67.54 & 62.33 & -5.21 &-7.7\%& 64.9 & -2.64 &-3.9\%& 68.1 & 0.56 &0.8\%\\
        \#20 & 70.24 & 63.61 & -6.63 & -9.4\%&67.75 & -2.49 &-3.5\%& 71.1 & 0.86 &1.2\%\\
        \#20 & 69.86 & 63.19 & -6.67 &-9.5\%& 67 & -2.86 &-4.1\%& 70.9 & 1.04 &1.5\%\\
        \#1 & 99.48  * & 98.95 & -0.53 &-0.5\%& 98.7 & -0.78 &-0.8\%& 99 & -0.48 &-0.5\% \\
        \#2 & 97.52  * & 97.88 & 0.36 &0.4\%& 95.4 & -2.12 &-2.2\% & 95.9 & -1.62 &-1.7\%\\
        \#3 & 39.74  * & 40.76 & 1.02 &2.6\%& 38.8 & -0.94&-2.4\% & 38.7 & -1.04 &-2.6\%\\
        \#4 & 35.63 & 36.27 & 0.63 &1.8\%& 36.2 & 0.56 &1.6\%& 36.2 & 0.56 &1.6\%\\
        \hline
        \quad & \quad&\multicolumn{3}{c|}{$\bar{\tau}=2.87$} & \multicolumn{3}{c|}{$\bar{\tau}=1.70$}
        & \multicolumn{3}{c|}{$\bar{\tau}=-0.115$}\\
        \quad & \quad&\multicolumn{3}{c|}{${s}=3.25$} & \multicolumn{3}{c|}{${s}=0.98$}
        & \multicolumn{3}{c|}{${s}=0.99$}\\
        \hline
        \end{tabular}
        }
         \caption{Estimated fixed loss from different approaches comparing with experimental data value, along with corresponding error $\tau$ and standard deviation $s$ ( *represents the upward escalator)}
         \label{T1}
    \end{table*}
    
    \subsection{Practical Comparison}
    We also compared the performance of the three approaches in four scenarios based on specific use cases commonly encountered in daily operation of escalators. 

\begin{figure*}
        \begin{adjustwidth}{-4cm}{-4cm}
        \centering
            \begin{subfigure}{0.30\linewidth}
                \centering
                \includegraphics[width=\linewidth]{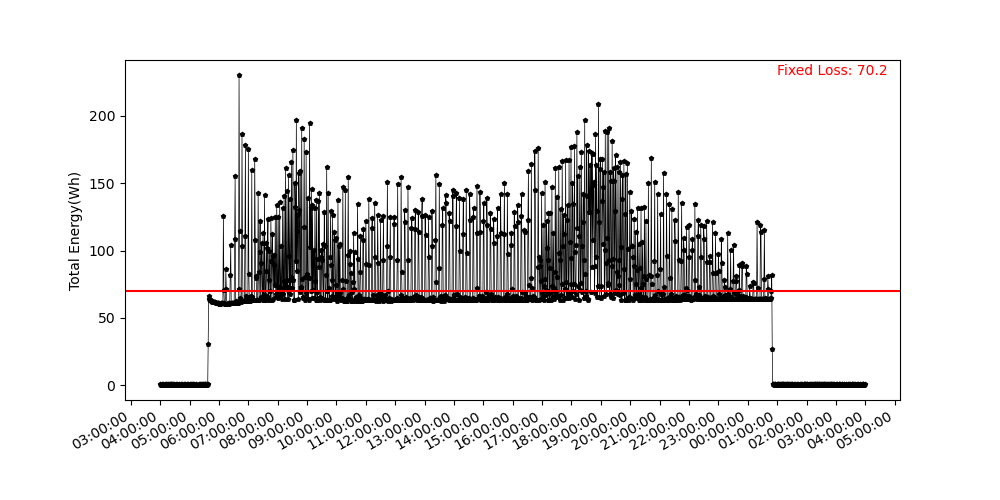}
                \caption{Classical Approach}
                \label{fig:comp_c_nvs}
            \end{subfigure}
            \begin{subfigure}{0.30\linewidth}
                \centering
                \includegraphics[width=\linewidth]{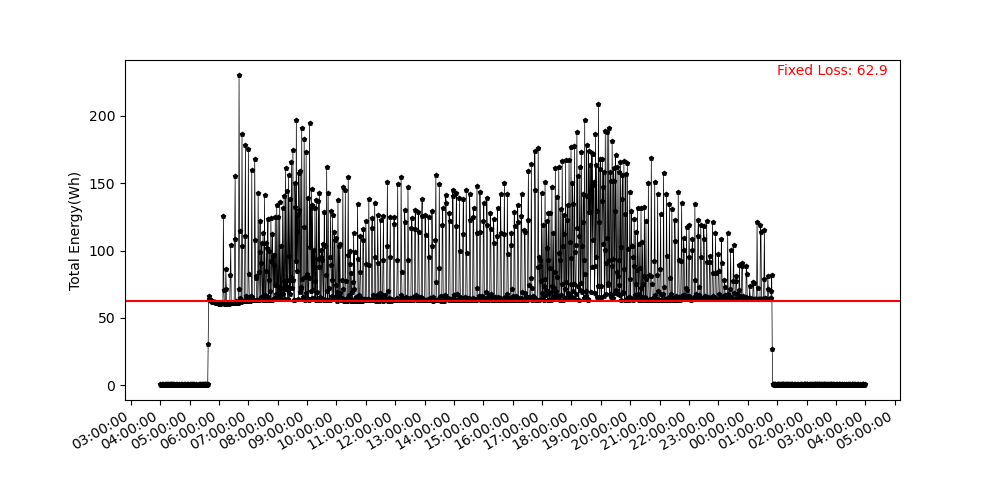}
                \caption{Engineering Approach}
                \label{fig:comp_m_nvs}
            \end{subfigure}
            \begin{subfigure}{0.30\linewidth}
                \centering
                \includegraphics[width=\linewidth]{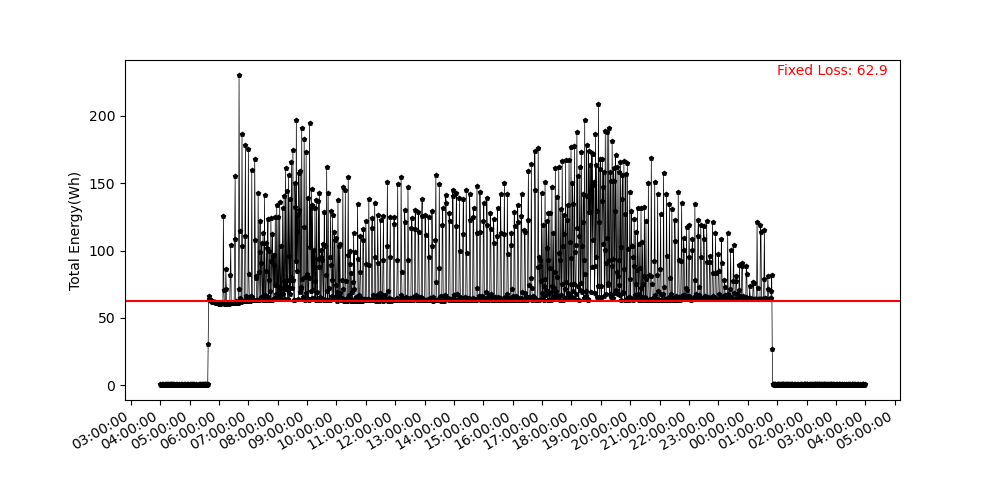}
                \caption{Optimization Approach}
                \label{fig:comp_opt_nvs}
            \end{subfigure}
        \end{adjustwidth}
        \caption{Fixed loss for the three approaches for escalators with last operating hour is non-vacant status (Escalator: \#23)}
        \label{fig:comp_nvs}
    \end{figure*}

    Firstly, we consider the scenario where passengers use the escalators until they are closed. This is common in Hong Kong where escalators and subway stations remain busy throughout their operating hours. Potentially this is a challenge for this classical approach which assumes the last 30 minutes are vacant or near-vacant. Figures \ref{fig:comp_nvs} shows a selected fixed loss energy profile and the corresponding fixed loss based on the three approaches. The selected profile is of an escalator with a busy last operating hour. It is evident that the classical approach is less reliable, in accurately estimating the fixed loss, while the engineering and optimization approaches are capable of handling this situation.

\begin{figure*}
        \begin{adjustwidth}{-4cm}{-4cm}
        \centering
            \begin{subfigure}{0.30\linewidth}
                \centering
                \includegraphics[width=\linewidth]{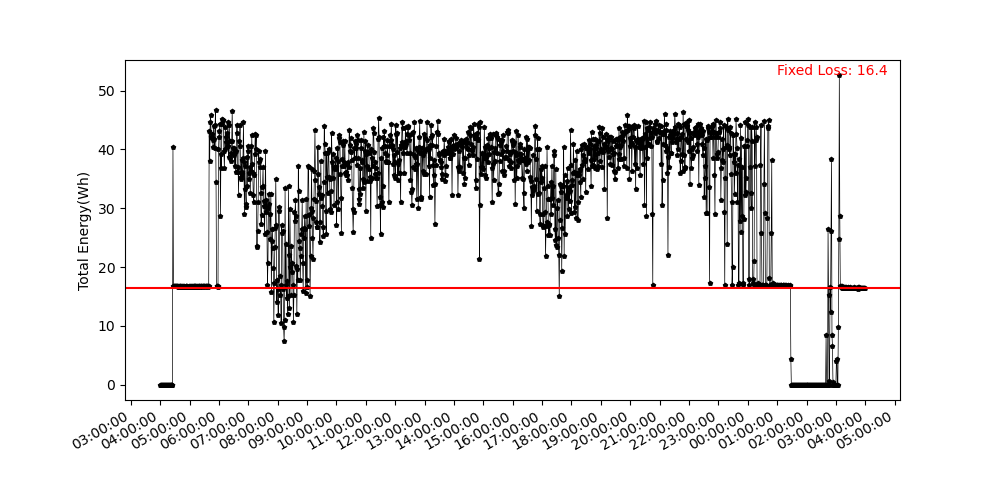}
                \caption{Classical Approach}
                \label{fig:comp_c_ess}
            \end{subfigure}
            \begin{subfigure}{0.30\linewidth}
                \centering
                \includegraphics[width=\linewidth]{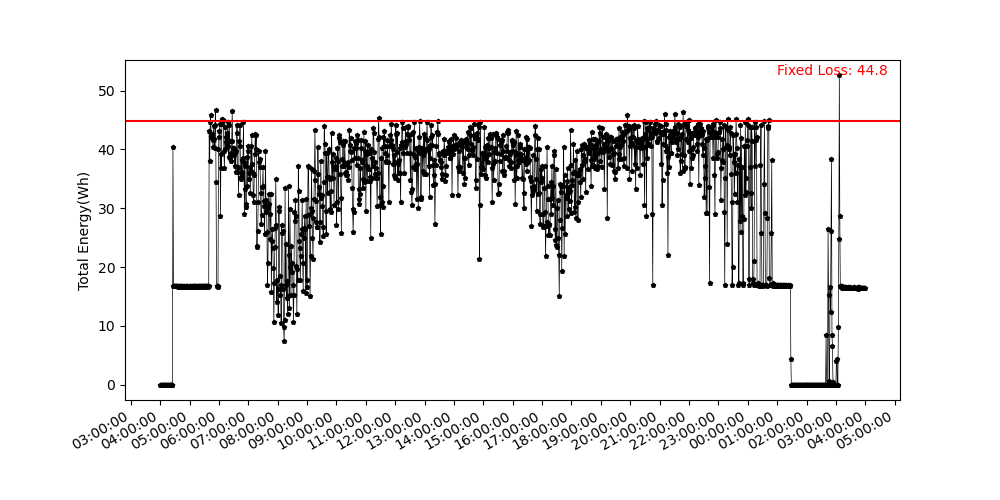}
                \caption{Engineering Approach}
                \label{fig:comp_m_ess}
            \end{subfigure}
            \begin{subfigure}{0.30\linewidth}
                \centering
                \includegraphics[width=\linewidth]{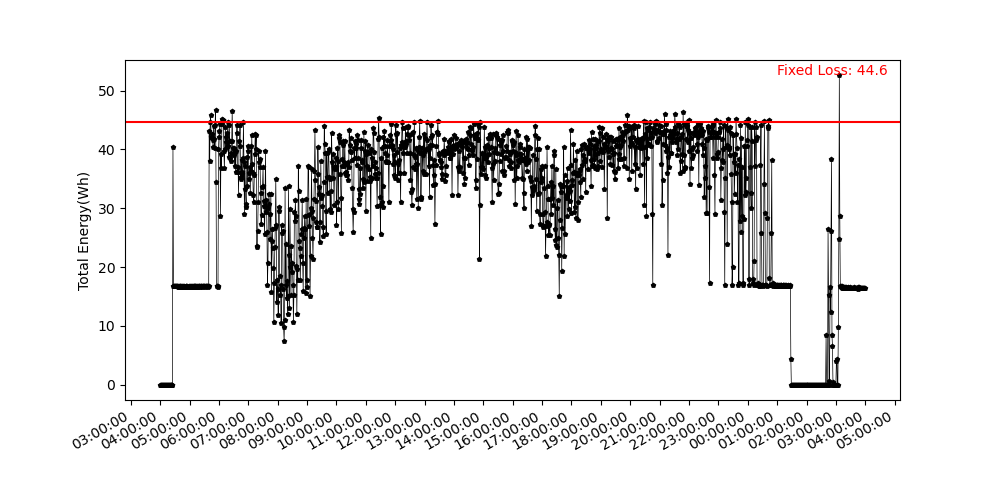}
                \caption{Optimization Approach}
                \label{fig:comp_opt_ess}
            \end{subfigure}
        \end{adjustwidth}
        \caption{Fixed loss for the three approaches for escalators with Energy-saving status (Escalator: \#7)}
        \label{fig:comp_ess}
    \end{figure*}

    Nowadays, modern escalators often come equipped with energy-saving functions that slow down or stop the escalator when no passengers are detected, resulting in lower energy consumption. This is illustrated in Figure \ref{fig:comp_ess}. The engineering and optimization approaches provide accurate estimates of the fixed loss under these conditions. However, in this case the classical approach estimates the fixed loss at the energy saving level as that was the escalator status in the last operating hour in this example.  
    
\begin{figure*}
        \begin{adjustwidth}{-4cm}{-4cm}
        \centering
            \begin{subfigure}{0.30\linewidth}
                \centering
                \includegraphics[width=\linewidth]{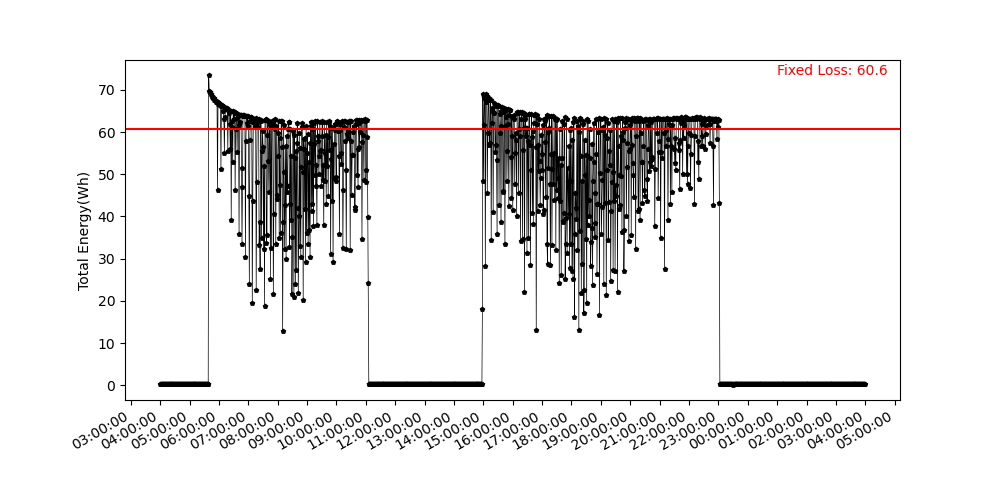}
                \caption{Classical Approach}
                \label{fig:comp_c_ms}
            \end{subfigure}
            \begin{subfigure}{0.30\linewidth}
                \centering
                \includegraphics[width=\linewidth]{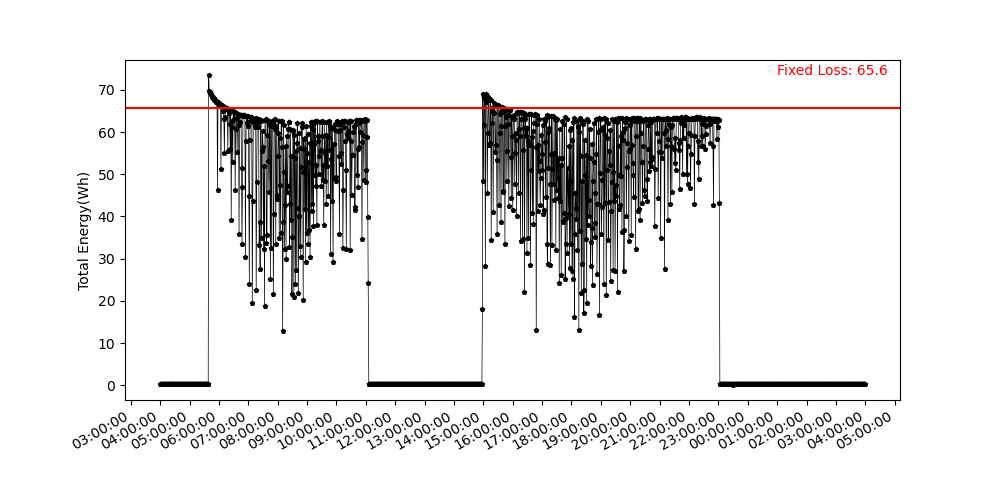}
                \caption{Engineering Approach}
                \label{fig:comp_m_ms}
            \end{subfigure}
            \begin{subfigure}{0.30\linewidth}
                \centering
                \includegraphics[width=\linewidth]{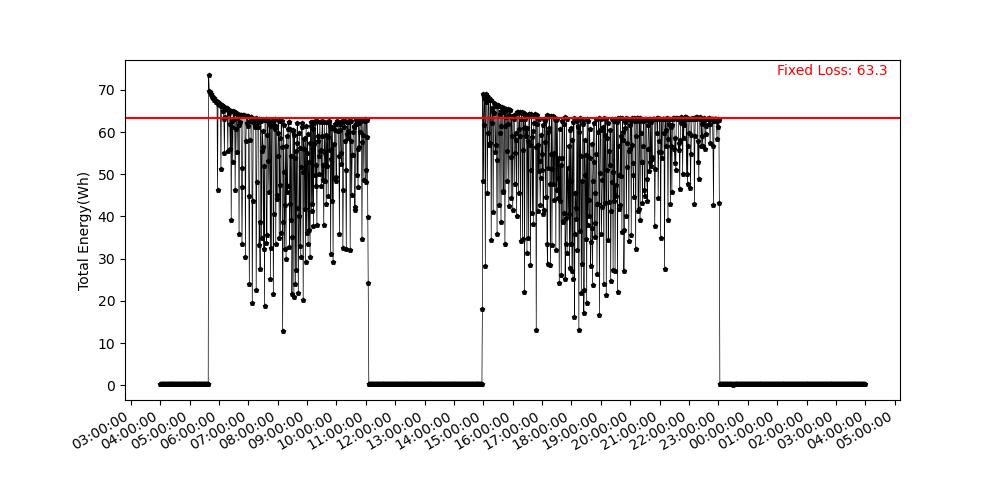}
                \caption{Optimization Approach}
                \label{fig:comp_opt_ms}
            \end{subfigure}
        \end{adjustwidth}
        \caption{Fixed loss for the three approaches for escalators with Multiple starting-ups in one day (Escalator: \#9)}
        \label{fig:comp_ms}
    \end{figure*}

    In some cases, escalators are manually turned off and restarted multiple times within a working day to conserve energy. However, due to the thermal process characteristics of the system, restarting the escalator uses more energy to warm up its components. This is illustrated in Figure \ref{fig:comp_ms}. From the result, the classical approach could not get the accurate value as the last operating hour is non-vacant. The engineering approach is biased towards the thermal process, leading to inaccurate estimates. The optimization approach provides a reasonable estimate in this scenario.
    
\begin{figure*}
        \begin{adjustwidth}{-4cm}{-4cm}
        \centering
            \begin{subfigure}{0.30\linewidth}
                \centering
                \includegraphics[width=\linewidth]{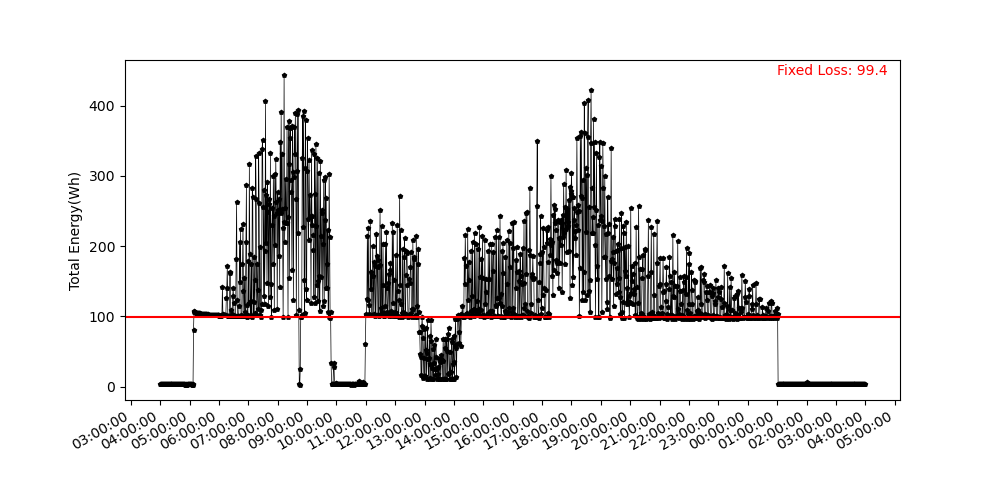}
                \caption{Classical Approach}
                \label{fig:comp_c_me}
            \end{subfigure}
            \begin{subfigure}{0.30\linewidth}
                \centering
                \includegraphics[width=\linewidth]{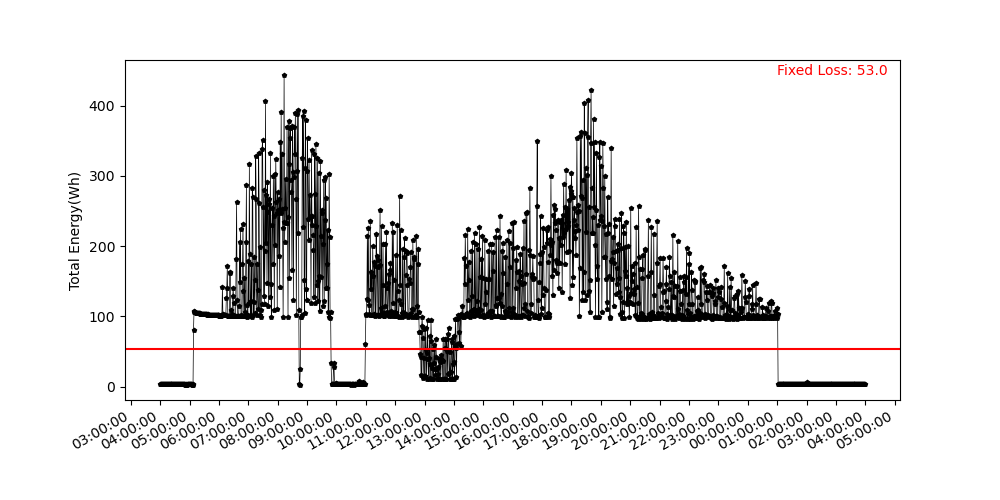}
                \caption{Engineering Approach}
                \label{fig:comp_m_me}
            \end{subfigure}
            \begin{subfigure}{0.30\linewidth}
                \centering
                \includegraphics[width=\linewidth]{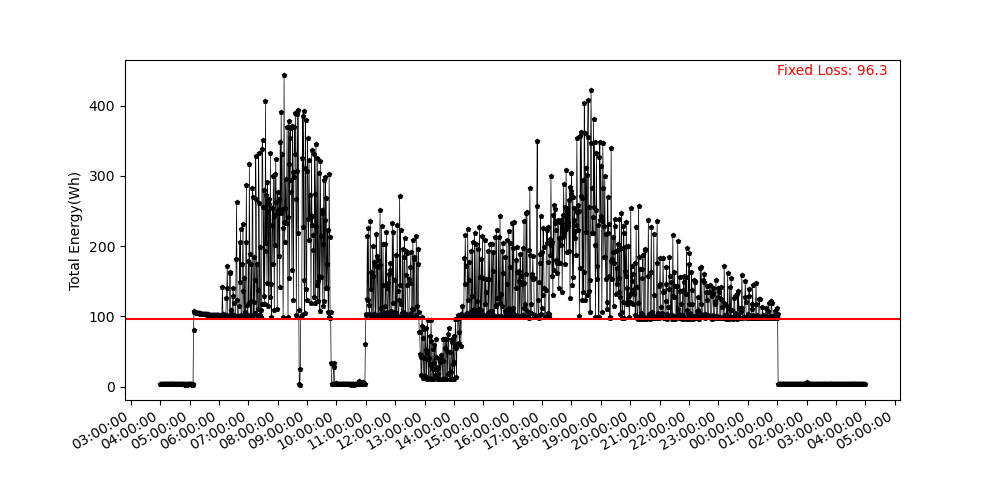}
                \caption{Optimization Approach}
                \label{fig:comp_opt_me}
            \end{subfigure}
        \end{adjustwidth}
        \caption{Fixed loss for the three approaches for escalators with a maintenance event (Escalator: \#1)}
        \label{fig:comp_me}
    \end{figure*}

    Regular maintenance events are essential to ensure the proper functioning of escalators. The energy consumption during maintenance event, as a result of the escalator being consistently turned on and off, is illustrated in Figure \ref{fig:comp_me}. The estimated fixed loss in Figure \ref{fig:comp_c_me} is influenced by the number of passengers, while the fixed loss with the engineering approach is largely affected by the energy consumption during maintenance time, as demonstrated in Figure \ref{fig:comp_m_me}. In contrast, the optimization approach provides a reliable and reasonable estimate even in the presence of maintenance events.
    
    To summarize, the classical approach is straightforward and easy to interpret but highly depends on the last operating hour being vacant, making it less robust. While the engineering approach is useful in most situations in normal daily usage, its highly biased estimate in maintenance or warming-up periods leads to it being less applicable in daily operation. The optimization approach provides superior accuracy and is suitable for all situations without requiring extensive data processing. Table \ref{table:compare} provides a summary of the strengths and weaknesses of each approach. Overall, the comparison of the performance of different methods shows that the optimization approach is the most accurate and robust for estimating time-varying fixed loss.

    \begin{table}[h]
    \begin{tabular}{ |p{3cm}|p{4.2cm}|p{4.2cm}|p{4.2cm} | }
    \hline
    \textbf{Methods}& \textbf{Classical} & \textbf{Engineering} & \textbf{Optimization} \\
    \hline
    \textbf{Description} & Average energy consumption in last operation hour. & Remove first operating hour, median of 5\% lowest (highest) energy consumption data points. & Within a threshold range, find a smallest (largest) value which most data points close to it.\\
    \hline
    \textbf{Pros}
    & 
    \begin{tabitem}
        \item Easy to understand and interpret
        \item No data processing required
    \end{tabitem}
    & 
    \begin{tabitem}
        \item Relatively easy to calculate
        \item Accurate in most daily normal operation
    \end{tabitem} 
    & 
    \begin{tabitem}
        \item Not restricted to operating hour, or passenger number
        \item No data processing needed
        \item Applicable to any specific operating condition
        \item Robust
    \end{tabitem}
    \tabularnewline
    \hline
    \textbf{Cons}
    & 
    \begin{tabitem}
        \item Highly dependent on operating time
        \item Heavily affected by specific operating conditions (maintenance and saving status etc.)
    \end{tabitem}
    & 
    \begin{tabitem}
        \item Dependent on operating time and location
        \item Extremely affected by specific operating condition (maintenance event, multiple start-ups etc.)
        \item Data processing is needed
    \end{tabitem} 
    & 
    \begin{tabitem}
        \item Computationally more intensive
    \end{tabitem}
    \tabularnewline
    \hline
    \end{tabular}
    \caption{Pros and Cons for the three approaches}
    \label{table:compare}
    \end{table}

\section{PHM Application: Abnormality Monitoring}
In this section, we discuss the concept of fixed loss energy as it relates to escalator PHM. Specifically, we investigate the monitoring of daily fixed loss energy for abnormal changes by using an Exponential Weighted Moving Average (EWMA). Use the optimization approach for daily fixed loss energy monitoring, setting a usage threshold of 0.3 as recommended in the previous section. The monitoring process is illustrated for two escalators.

\subsection{Methodology}
Exponential weighted moving average (EWMA) control charts were firstly proposed by Roberts \cite{Roberts59} as a statistical process control method for detecting small shifts in the process mean using a weighted average of all available historical data. EWMA’s recursive property leads to exponentially decaying weights. We constructed a dynamic EWMA process with 30 days window. The EWMA statistics $Z_t$ for day $t$ is:

\begin{equation}
Z_t = \lambda F_t + (1-\lambda) Z_{t-1}.
\end{equation} 
The constant $\lambda$ represents the weight of the previous data set considered in the calculation of the current EWMA statistic where $0<\lambda<1$, $F_{t}$ is the fixed loss at day $t$. The chart signals when $\lvert Z_k \rvert\ge k$. 

The fixed loss fluctuates over time due to varying operating condition (refer to Figure \ref{EWMA}). We are interested in sudden (large) shift in the mean. To accommodate for gradual changes, we apply moving window based control limits and set te window to 30 days. 


The time-varying upper and lower EWMA control limits for the $Z_t$ are:
\begin{equation}
\begin{aligned}
UCL = \mu_w +k\sigma_{EWMA} = \mu_w +k\sqrt{\frac{\lambda}{2-\lambda}\sigma_w^2},\\
LCL = \mu_w - k\sigma_{EWMA} = \mu_w -k\sqrt{\frac{\lambda}{2-\lambda}\sigma_w^2}
\end{aligned}
\end{equation}
Where $\mu_w$ and $\sigma_w$ are the mean and standard deviation that we compute based on the past 30 days (a moving window). In order to have robust estimator we apply the sample trimean, as advised in Zwetsloot et. al., \cite{Zwetsloot16}. It is defined as 
\begin{equation}
    \mu_w = \frac{Q_{1,w}+2Q_{2,w}+Q_{3,w}}{4}
\end{equation}
where $Q_{1,w}$ is the 25\% quantile of the past 30 fixed loss and $Q_{2,w}$ is the median and $Q_{3,w}$ is the 75\% quantile. For $\sigma_w^2$ we also use a robust estimator, the Interquantile range: 

\begin{equation}
\sigma_w^2 = \frac{Q_{3,w}-Q_{1,w}}{d_w}
\end{equation}
where $d_w=0.779$ is the unbiasing constant. 

The study conducted by Lucus and Saccucci \cite{Lucus12} analyzed the properties of EWMA and presented recommendations for selecting appropriate values of $\lambda$ and $k$. Their study suggests 0.25 for $\lambda$. Also, we set $k=2.924$ for the in-control ARL to be approximately 400.

\subsection{Numerical Example}
\begin{figure}[h!]
        \centering
        \begin{subfigure}[b]{1\textwidth}
            \centering
            \includegraphics[scale=0.45]{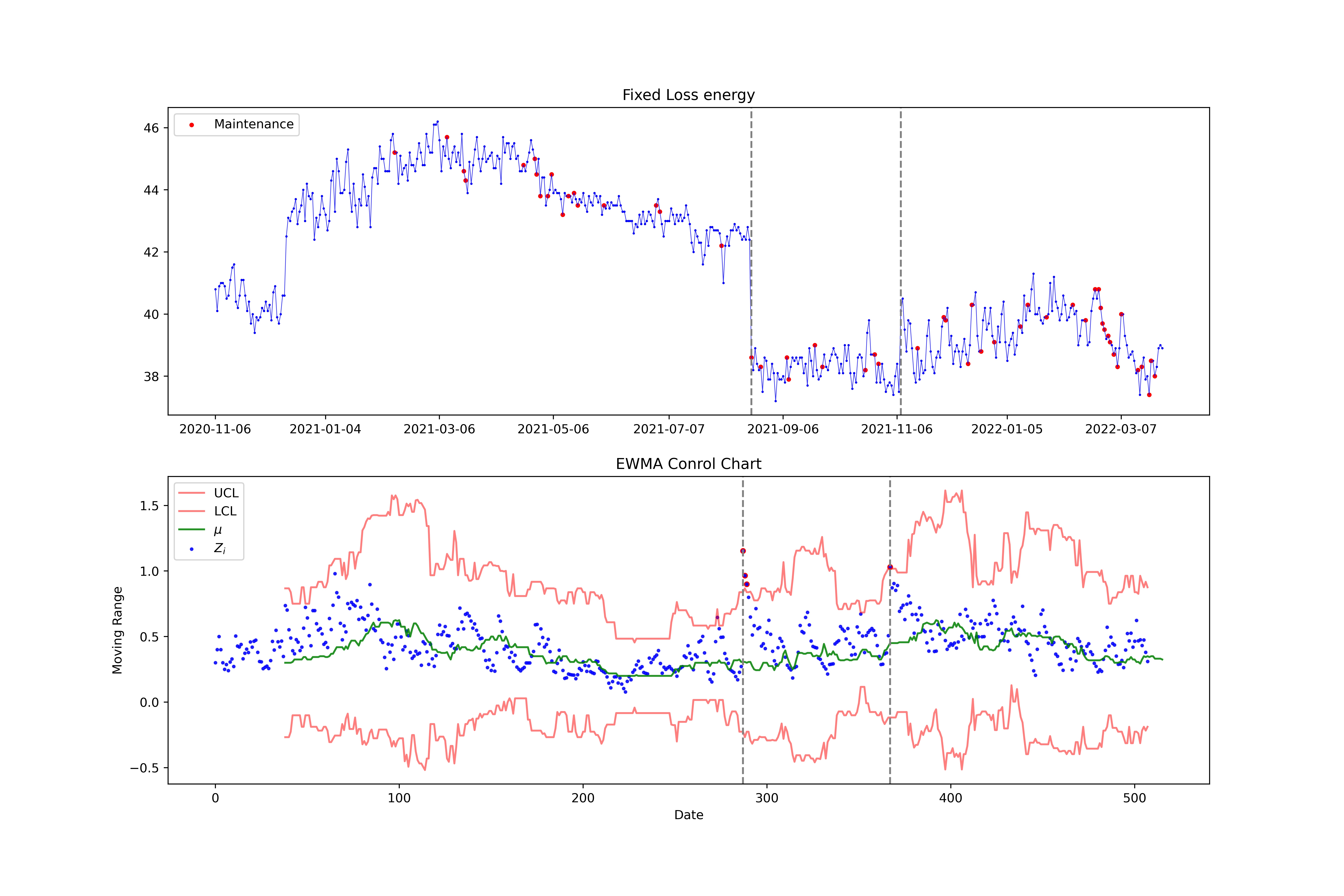}
            \caption{Escalator: \#3 (Upward)}
            \label{EWMA:upward}
        \end{subfigure}\\
        
        \begin{subfigure}[b]{1\textwidth}
            \centering
            \includegraphics[scale=0.45]{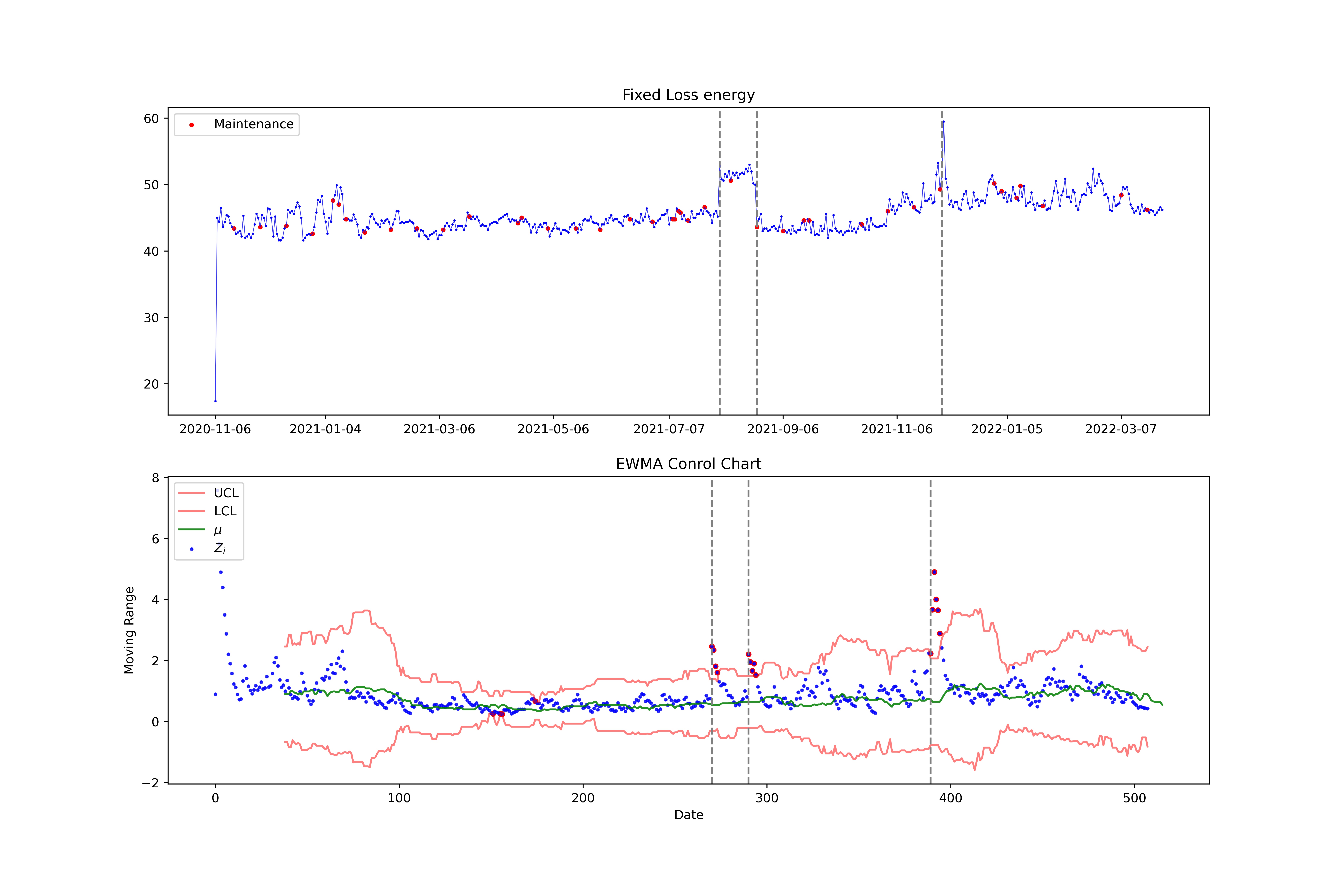}
            \caption{Escalator: \#7 (Downward)}
            \label{EWMA:downward}
        \end{subfigure}
        \caption{EWMA control chart for upward and downward escalators}
        \label{EWMA}
\end{figure}

In this section, we monitor the fixed loss of two escalators using the EWMA chart as introduced in Section 5.1. We calculated the fixed loss energy for a period of 400 days by using the optimization approach. An EWMA control chart was generated based on the last 30 days of data for each day. Figure \ref{EWMA} shows both fixed loss over time and corresponding control chart. The majority of the resulting statistics fell within the control limit, indicating that fixed loss energy consumption was regular during this time period. We annotated all data points out of the control limit with grey lines in both fixed loss energy and EWMA chart. Figure \ref{EWMA:upward} illustrates fluctuations in the fixed loss, and the monitoring process successfully detected sudden changes. Notably, at the beginning of the period, there was a substantial increase in fixed loss. However, due to the 30-day moving window used for the EWMA chart, it took some time for this initial surge to be captured. In Figure \ref{EWMA:downward}, the fixed loss remained relatively stable, with minor fluctuations around an average value of approximately 45 Wh. Nevertheless, significant changes were observed during the middle period, indicating a departure from the usual pattern.

Furthermore, we incorporated maintenance records provided by the escalator operator into the analysis, marking them in red on the original fixed loss energy graph. It is worth noting that there was a drop in fixed loss energy after some maintenance events. This is evident in the first abnormal data point in Figure \ref{EWMA:upward} and the second grey line in Figure \ref{EWMA:downward}. This observation emphasizes the impact of maintenance activities on the energy consumption. It indicates that proper maintenance can lead to improved energy efficiency and highlights the importance of considering maintenance practices in optimizing of escalator performance and energy saving. 

\section{Conclusion Remarks}
The ability to collect data and its storage has significantly improved, does allowing condition monitoring. Prognostic health management (PHM) is becoming increasingly important in both industry and academia, as it allows for more proactive and cost-effective maintenance. The energy consumption of an escalator is related to its mechanical design and health status, making it a suitable application for PHM processes.

In this study, we analyzed the energy consumption of escalators using a large dataset collected over a 20-month period. We introduced a novel approach, the optimization approach, for accurately estimating the fixed loss energy of escalators. Additionally, we conducted the experiments to fine-tune parameters and compare all approaches in different scenarios commonly occurring in our daily operation. The results demonstrate that the optimization approach is highly accurate and robust, making it applicable to a wide range of scenarios. In contrast, the classical approach and the engineering approach have limitations in different common daily operating scenarios.

The optimization approach has the potential to significantly improve energy efficiency and reduce energy costs in escalator operation. By effectively managing and monitoring the fixed loss, improvements can be made to the overall performance and energy consumption of escalators. This study contributes to the development of efficient Prognostic Health Management (PHM) processes for escalators.

A limitation of this study is that it only considers escalators in Hong Kong that are uni-directional. For bi-directional escalators, modified solutions need to be developed that can take into account the fact that fixed loss depends on the direction of the escalator.

\section*{Data availability statement}
The data that support the findings of this study are available from MTR Corporation Ltd. Restrictions apply to the availability of these data, which were used under license for this study. Data are available through Inez Zwetsloot subject to the permission of MTR Corporation Ltd. 

\section*{Acknowledgment}

The work in this paper was supported by a grant from the Innovation and Technology Fund (ref: PRP-008-20FX) and MTR Corporation Limited.


\begin{thebibliography}{99}
\bibitem{Tsui15}
Tsui, K. L.,  Chen, N., Zhou, Q., Hai, Y., \& Wang, W.  (2015). Prognostics and health management: A review on data driven approaches. \emph{Mathematical Problems in Engineering, 2015}.

\bibitem{Batzel09}
Batzel, T. D.,  \& Swanson, D. C. (2009). Prognostic health management of aircraft power generators. \emph{IEEE Transactions on Aerospace and electronic systems, 45(2)}, 473-482.

\bibitem{Che19}
Che, C., Wang, H., Fu, Q., \& Ni, X. (2019). Combining multiple deep learning algorithms for prognostic and health management of aircraft. \emph{Aerospace Science and Technology, 94}, 105423.

\bibitem{Hu22}
Hu, Y., Miao, X., Si, Y., Pan, E., \& Zio, E. (2022). Prognostics and health management: A review from the perspectives of design, development and decision. \emph{Reliability Engineering \& System Safety, 217}, 108063.

\bibitem{Jonge17}
de Jonge, B., Teunter, R., \^ Tinga, T.  (2017). The influence of practical factors on the benefits of condition-based maintenance over time-based maintenance. \emph{Reliability engineering \& system safety, 158}, 21-30.

\bibitem{Dimitris22}
Kampitsis, D., \& Panagiotidou, S.  (2022). A Bayesian condition-based maintenance and monitoring policy with variable sampling intervals. \emph{Reliability Engineering \& System Safety, 218}, 108159.

\bibitem{Zheng21}
Zheng, M., Ye, H., Wang, D., \& Pan, E. (2021). Joint optimization of condition-based maintenance and spare parts orders for multi-unit systems with dual sourcing. \emph{Reliability engineering \& system safety, 210}, 107512.

\bibitem{Liu22}
Liu, Q., Ma, L., Wang, N., Chen, A., \& Jiang, Q. (2022). A condition-based maintenance model considering multiple maintenance effects on the dependent failure processes. \emph{Reliability Engineering \& System Safety, 220}, 108267.

\bibitem{Xing20}
Xing, Y., Chen, S., Zhu, S., \& Lu, J. (2020). Analysis factors that influence escalator-related injuries in metro stations based on bayesian networks: A case study in China. \emph{International journal of environmental research and public health, 17(2)}, 481. 

\bibitem{Al-sharif11}
Al-Sharif, L. (2011). Modelling of escalator energy consumption. \emph{Energy and Buildings, 43(6)}, 1382-1391.

\bibitem{Uimonen17}
Uimonen, S., Tukiam, T., Siikonen, M. L., \& Lehtonen, M. (2017). Predicting the annual escalator energy consumption based on short-term measurements. \emph{Journal of Building Engineering, 13}, 319-325.

\bibitem{De12}
De Almeida, A., Hirzel, S., Patrão, C., Fong, J., \& Dütschke, E. (2012). Energy-efficient elevators and escalators in Europe: An analysis of energy efficiency potentials and policy measures. \emph{Energy and buildings, 47},  151-158.

\bibitem{Uimonen20}
Uimonen, S., Tukia, T., Ekström, J., Siikonen, M. L., \& Lehtonen, M. (2020). A machine learning approach to modelling escalator demand response.  \emph{Engineering Applications of Artificial Intelligence, 90}, 103521.


\bibitem{Elasha14}
Elasha, F., Ruiz-Cárcel, C., Mba, D., Kiat, G., Nze, I., \& Yebra, G. Pitting detection in worm gearboxes with vibration analysis. \emph{Engineering Failure Analysis, 42}, 366-376. 

\bibitem{Huo20}
Huo, M., Li, X., Wei, G., \& Zhao, C.  (2019, October). Application research of escalators status monitor and forecast based on vibration analysis. In \emph{International Conference on Electrical and Information Technologies for Rail Transportation} (pp. 419-428). Singapore: Springer Singapore. 

\bibitem{Ma09}
Ma, W. W., Liu, X. Y., Li, L., Shi, X., \& Zhou, C. Q. (2009). Research on the waiting time of passengers and escalator energy consumption at the railway station. \emph{Energy and Buildings, 41(12)}, 1313-1318.

\bibitem{Carrillo13}
Carrillo, C., Díaz-Dorado, E., Cidrás, J., \& Silva-Ucha, M. (2013). A methodology for energy analysis of escalators. \emph{ Energy and buildings, 61}, 21-30.

\bibitem{Al-sharif98}
Al-Sharif, L. (1998). The general theory of escalator energy consumption with calculations and examples. \emph{Elevator World, 46}, 74-79.

\bibitem{Kuutti13}
Kuutti, J., Sepponen, R. E., \& Saarikko, P. (2013, September). Escalator power consumption compared to pedestrian counting data. In \emph{2013 International Conference on Applied Electronics}(pp. 1-4). IEEE.

\bibitem{Roberts59}
Roberts, S. W. (2000). Control Chart Tests Based on Geometric Moving Averages. \emph{Technometrics, 42(1)}, 97-101. 

\bibitem{Zwetsloot16}
Zwetsloot, I. M., Schoonhoven, M., \& Does, R. J. (2016). Robust point location estimators for the EWMA control chart. \emph{Quality Technology \& Quantitative Management, 13(1)}, 29-38.












\bibitem{Lucus12}
Lucas, J. M., \& Saccucci, M. S. (1990). Exponentially weighted moving average control schemes: properties and enhancements." \emph{Technometrics, 32(1)}, 1-12.
\end{thebibliography}
\end{document}